\documentclass[twocolumn]{aastex63}

\newcommand\Msun{M$_\odot$}
\newcommand\oi{[O\,I]}
\newcommand\caii{[Ca\,II]}
\graphicspath{{./}{figures/}}
\begin{document}

\submitjournal{ApJL}

\title{The peculiar Ca-rich SN\,2019ehk: Evidence for a Type IIb core-collapse supernova from a low mass stripped progenitor}

\correspondingauthor{Kishalay De}
\email{kde@astro.caltech.edu}

\author{Kishalay De}
\affil{Cahill Center for Astrophysics, California Institute of Technology, 1200 E. California Blvd. Pasadena, CA 91125, USA}

\author{U. Christoffer Fremling}
\affil{Cahill Center for Astrophysics, California Institute of Technology, 1200 E. California Blvd. Pasadena, CA 91125, USA}

\author{Avishay Gal-Yam}
\affil{Benoziyo Center for Astrophysics, The Weizmann Institute of Science, Rehovot 76100, Israel}

\author{Ofer Yaron}
\affil{Benoziyo Center for Astrophysics, The Weizmann Institute of Science, Rehovot 76100, Israel}

\author{Mansi M. Kasliwal}
\affil{Cahill Center for Astrophysics, California Institute of Technology, 1200 E. California Blvd. Pasadena, CA 91125, USA}

\author{S. R. Kulkarni}
\affil{Cahill Center for Astrophysics, California Institute of Technology, 1200 E. California Blvd. Pasadena, CA 91125, USA}

\begin{abstract}
The nature of the peculiar `Ca-rich' SN\,2019ehk in the nearby galaxy M100 remains unclear. Its origin has been debated as either a stripped core-collapse supernova or a thermonuclear helium detonation event. Here, we present very late-time photometry of the transient obtained with the Keck I telescope at $\approx280$\,days from peak light. Using the photometry to perform accurate flux calibration of a contemporaneous nebular phase spectrum, we measure an \oi\ luminosity of $(0.19-1.08)\times10^{38}$\,erg\,s$^{-1}$ and \caii\ luminosity of $(2.7-15.6)\times10^{38}$\,erg\,s$^{-1}$ over the range of the uncertain extinction along the line of sight and distance to the host galaxy. We use these measurements to derive lower limits on the synthesized oxygen mass of $\approx0.004-0.069$\,\Msun. The oxygen mass is a sensitive tracer of the progenitor mass for core-collapse supernovae, and our estimate is consistent with explosions of very low mass CO cores of $1.45-1.5$\,\Msun, corresponding to He core masses of $\approx1.8-2.0$\,\Msun. We present high quality peak light optical spectra of the transient and highlight features of hydrogen in both the early (`flash') and photospheric phase spectra, that suggest the presence of $\gtrsim0.02$\,\Msun\ of hydrogen in the progenitor at the time of explosion. The presence of H, together with the large \caii/\oi\ ratio ($\approx10-15$) in the nebular phase is consistent with SN\,2019ehk being a Type\,IIb core-collapse supernova from a stripped low mass ($\approx9-9.5$\,\Msun) progenitor, similar to the Ca-rich SN\,IIb iPTF\,15eqv. These results provide evidence for a likely class of `Ca-rich' core-collapse supernovae from stripped low mass progenitors in star forming environments, distinct from the thermonuclear Ca-rich gap transients in old environments.
\end{abstract}


\section{Introduction}
Ca-rich gap transients are an intriguing class of faint and fast evolving explosions characterized by their conspicuous strong \caii\ $\lambda\lambda 7291, 7324$ emission (compared to \oi\ $\lambda\lambda 6300, 6364$) in the nebular phase \citep{Fillipenko2003, Perets2010, Kasliwal2012a, Valenti2014, Lunnan2017, Gal-Yam2017, Milisavljevic2017, De2020}. Tracking down the progenitors and explosion mechanisms of these unique transients is important for our understanding of the fates of close binary systems, the progenitors of Type Ia SNe and the cosmic nucleosynthesis of Ca \citep{Mulchaey2014, Frohmaier2019, De2020}.

The peculiar SN\,2019ehk was discovered in the galaxy M100 \citep{Grzegorzek2019}, and subsequent follow-up showed that the source exhibited fast photometric and spectroscopic evolution to the nebular phase dominated by strong \caii\ emission, consistent with several known properties of Ca-rich events \citep{Jacobson-Galan2020, Nakaoka2020}. \citet{Jacobson-Galan2020} suggested that its early fast spectroscopic evolution and double peaked light curve is likely explained with an explosive thermonuclear detonation ignited during a white dwarf merger involving a low mass hybrid white dwarf. However, archival Hubble Space Telescope (HST) images could not rule out a core-collapse explosion from a $<10$\,\Msun\ massive star. On the other hand, \citet{Nakaoka2020} favored a scenario involving a low-mass core-collapse supernova (SN) from an inflated and `ultra-stripped' He star in a close binary system \citep{Tauris2013,Tauris2015} -- a channel which has been suggested to lead to the formation of neutron stars in compact binary systems. As one of the nearest potential members of the class of Ca-rich events, constraining the nature of the progenitor of SN\,2019ehk can reveal important clues to the broader population of events.

With the advent of large systematic experiments for supernova (SN) classification, it is now well established that Ca-rich gap transients are relatively common ($\approx 15$\% of the SN\,Ia rate) and predominantly occur in old environments in the outskirts of early type galaxies, suggesting progenitor systems likely involving explosive He shell burning on low mass white dwarfs \citep{Perets2010, Kasliwal2012a, Lunnan2017, Frohmaier2018, De2020}. The dominance of cooling via \caii\ emission as opposed to Fe emission (seen in normal Type Ia SNe) has been recently shown to be a hallmark feature of explosions involving shell detonations \citep{Waldman2011, Dessart2015a} with low total (core + shell) masses \citep{Polin2019b}.

However, the discovery of Ca-rich SNe such as iPTF\,15eqv \citep{Milisavljevic2017} and iPTF\,16hgs \citep{De2018b} in actively star forming environments (as in the case of SN\,2019ehk) have also led to suggestions involving core-collapse supernovae from low mass progenitors. Yet, the high \caii/\oi\ ratio seen in the population of Ca-rich transients \citep{Valenti2014, Milisavljevic2017, De2020} is strikingly different from that seen in normal stripped core-collapse SNe \citep{Fang2019}. 

Oxygen in the ejecta of core-collapse SNe is formed primarily in the hydrostatic burning phase of the progenitor (increasing with zero age main sequence mass), while Ca is explosively synthesized by O burning \citep{Fransson1989, Woosley2007}. As a result, the O mass in the ejecta and Ca/O ratio is a powerful tracer of the progenitor mass for core-collapse SNe \citep{Fransson1989, Jerkstrand2014, Jerkstrand2015}. In the case of the Ca-rich SN\,2005cz, \citet{Kawabata2010} thus first suggested that the high \caii/\oi\ ratio could be explained by an explosion of a low mass progenitor that was stripped by a binary companion.

In this paper, we attempt to constrain the progenitor of SN\,2019ehk with new late-time photometry and high quality optical spectra obtained near peak light. Section \ref{sec:observations} provides an overview of the observations and data analysis procedures. We use the observations to constrain the composition of the ejecta in both the early photospheric and late nebular phase in Section \ref{sec:results}. We present a discussion on the likely progenitor for SN\,2019ehk in Section \ref{sec:discussion} and conclude with a summary in Section \ref{sec:summary}. We adopt a nominal distance of 16.2\,Mpc and redshift of $z = 0.005$ to M100 for the rest of this work \citep{Folatelli2010}. However, there is a span of $\approx 14.2 - 21.4$\,Mpc in reported distances using Cepheid variables (e.g. \citealt{Freedman2001}; as in the NASA Extragalactic Database), which we use as the range of possible distances to the host galaxy in estimating uncertainty intervals.

\section{Observations and Analysis}
\label{sec:observations}

\begin{figure}[!ht]
    \centering
    \includegraphics[width=0.49\textwidth]{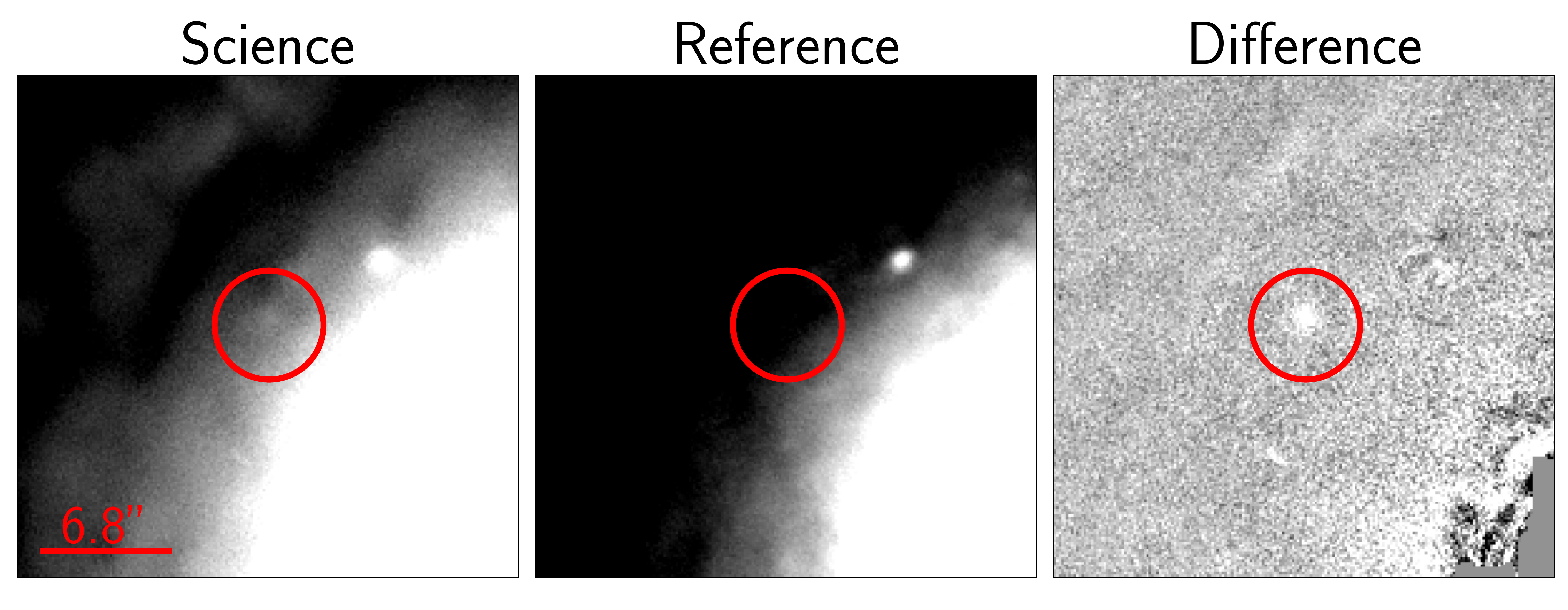}
    \includegraphics[width=0.49\textwidth]{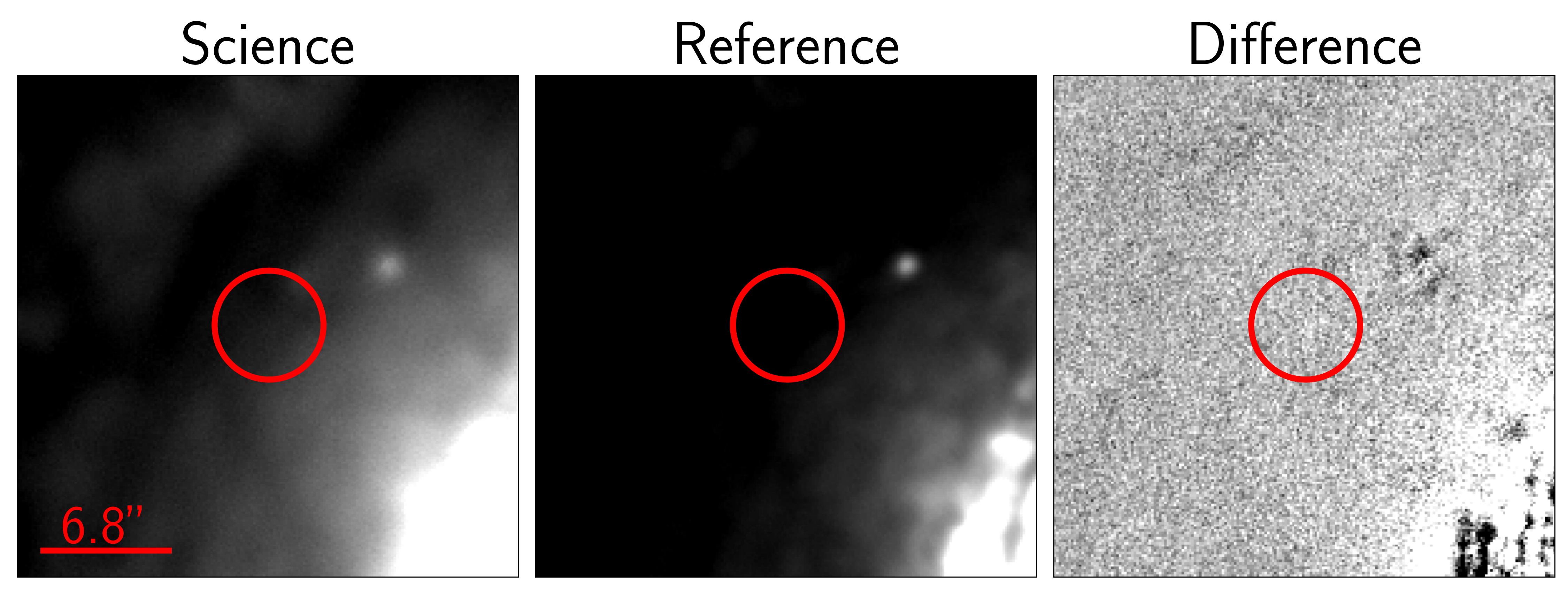}
    \caption{(Top panel) Late-time $I$-band detection of SN\,2019ehk with the Keck-I telescope, with North up and East left. The left panel is the image taken at $\approx 280$ days from peak light, the middle panel is the template image acquired at $\approx 400$ days after peak light and the right panel is the difference image obtained after image subtraction. (Bottom panel) Same as top panel showing non-detection of SN\,2019ehk at the same epoch in $g$ band.}
    \label{fig:subtraction}
\end{figure}

We obtained one epoch of late-time imaging of SN\,2019ehk with the Low Resolution Imaging Spectrometer (LRIS; \citealt{Oke1995}) on the Keck-I telescope on UT 2020-02-18.62, at a phase of $\approx 280$\,days from $r$-band peak, for a total exposure time of 300\,s and 390\,s in $g$ and $I$ bands respectively. We obtained a reference epoch for the source on UT 2020-06-23.32 to use as a template for image subtraction of the host galaxy light, for a total exposure time of 520\,s and 440\,s in $g$ and $I$ bands respectively. The data were reduced using \texttt{lpipe} \citep{Perley2019} and image subtraction was performed using \texttt{HOTPANTS} \citep{Becker2015}.

Photometric calibration was performed against SDSS catalog magnitudes of secondary standards in the field. The source is clearly detected in the $I$-band subtracted image at a magnitude of $I = 22.10 \pm 0.15$\,AB\,mag (Figure \ref{fig:subtraction}), while the source is not detected in $g$-band to a $3\sigma$ depth of 23.55\,AB\,mag. Based on the observed decay rate of the late-time light curve ($\gtrsim 0.02$\,mag\,day$^{-1}$; \citealt{Jacobson-Galan2020}), we expect the flux of the source at the template image epoch ($\approx 400$\,days) to be $\gtrsim 10\times$ smaller than the science epoch, and thus not contaminate our measurements significantly.

We use the observed late-time photometry to calibrate the published late-time spectrum at $\approx 260$ days in \citet{Jacobson-Galan2020}, noting that the strong \caii\ line falls completely within the observed $I$ band. We perform spectrophotometric calibration by convolving the filter function with the observed spectrum, and then measure the resulting line fluxes by trapezoidal integration of the respective wavelength regions. Uncertainties in this method are estimated by Monte Carlo sampling of the estimated fluxes by adding noise (scaled to nearby regions with no line emission) to the line profile, and add it in quadrature to the uncertainty of the photometric measurement. We measure the resulting \caii\ line flux to be $(4.0 \pm 0.6) \times 10^{-15}$\,erg\,cm$^{-2}$\,s$^{-1}$ and the corresponding observed \oi\ line flux to be $(2.1 \pm 0.4) \times 10^{-16}$\,erg\,cm$^{-2}$\,s$^{-1}$. 

We also present optical spectroscopy of the transient obtained with the Double Beam Spectrograph (DBSP; \citealt{Oke1982}) on the Palomar 200-inch telescope (P200) on UT 2019-05-13, corresponding to a phase of $\approx +0$\,days from $r$-band peak. The DBSP data were reduced using the \texttt{pyraf-dbsp} pipeline \citep{Bellm2016}. The data presented here will be publicly released on WISERep \citep{Yaron2012}.

\section{Results}
\label{sec:results}

\subsection{Constraints on host galaxy extinction}

There is evidence for significant host galaxy extinction towards SN\,2019ehk \citep{Jacobson-Galan2020, Nakaoka2020}. A deep Na I D line is clearly detected in its peak light spectra, and suggest a large host extinction of $E(B-V) \gtrsim 1$ based on canonical relationships between between $E(B-V)$ and the equivalent width ($EW$) of the Na D line \citep{Poznanski2012}. However, the very large equivalent width ($EW \approx 3$\,\AA) falls in a regime where published relationships become uncertain \citep{Poznanski2012}. The adopted extinction thus introduces an additional uncertainty in the determination of the absolute luminosity of the supernova and the nebular phase spectral lines.

However, the double-peaked light curve of SN\,2019ehk shares several similarities with previously reported fast evolving Type I SNe in the literature, including the SN\,Ic iPTF\,14gqr \citep{De2018a} as well as the SN\,Ib iPTF\,16hgs \citep{De2018b}. \citet{Nakaoka2020} show that the photometric properties of SN\,2019ehk can match either the low peak luminosity of iPTF\,16hgs or the higher luminosity of iPTF\,14gqr for assumed extinctions of $E(B-V) = 0.5$\,mag and $E(B-V) = 1.0$\,mag respectively, with the true value being likely in between these two\footnote{The value of $E(B-V) = 0.47$ adopted in \citet{Jacobson-Galan2020} is at the lower limit of the range of extinction assumed here}. Taking these two values of extinction as limiting cases, we obtain extinction corrected \oi\ flux of $(7.2 \pm 1.3) \times 10^{-16}$\,erg\,cm$^{-2}$\,s$^{-1}$ and $(2.5 \pm 0.4) \times 10^{-15}$\,erg\,cm$^{-2}$\,s$^{-1}$ respectively assuming $R_V = 3.1$ and a \citet{Cardelli1989} extinction law. The corresponding \caii\ line fluxes are $(1.1 \pm 0.2) \times 10^{-14}$\,erg\,cm$^{-2}$\,s$^{-1}$ and $(2.9 \pm 0.4) \times 10^{-14}$\,erg\,cm$^{-2}$\,s$^{-1}$. 

\subsection{Constraints on the oxygen mass}

\citet{Uomoto1986} provides an analytical formula to calculate the minimum O mass required for a given \oi\ luminosity, which depends on the temperature of the emitting region. The relationship holds in the high density limit ($N_e \gtrsim 10^6$\,cm$^{-3}$) where the electron density is above the \oi\ critical density ($\sim 7 \times 10^5$\,cm$^{-3}$), and is estimated to hold in this case for the estimated ejecta mass of $\approx 0.5$\,\Msun\ \citep{Jacobson-Galan2020, Nakaoka2020}. However, we caution that such O mass estimates assume that the radioactive power deposited in the O-rich shells of the ejecta is released via cooling in the \oi\ lines. \citet{Dessart2020} show that even small amounts of Ca mixing ($\sim 0.01$ by mass fraction) from the underlying Si-rich layers can drastically reduce the \oi\ line fluxes since \caii\ is a much more effective coolant than \oi. In the case of SN\,2019ehk, it is clear that the majority of the cooling is arising from the \caii\ line, which may be due to either a very low O layer mass compared to the Si-rich layer, or due to enhanced mixing of Ca into O-rich layers. As such, these O mass estimates should be treated as lower limits on the O mass in the ejecta.

While the temperature can be constrained with the line ratio of the \oi\ $\lambda 5577$\,\AA\, line to the \oi\ $\lambda\lambda 6300, 6364$ doublet \citep{Houck1996}, the weak \oi\ line in the SN\,2019ehk spectrum at $+260$\,days does not allow this measurement. Instead, we adopt a range of typical values estimated from the \oi\ emission in other core-collapse SNe of $\approx 3400 - 4000$\,K \citep{Sollerman1998, Elmhamdi2011}. We derive lower limits on the O mass in the range of $\approx 0.004 - 0.069$\,M$_\odot$ over the range of temperature, extinction and distance estimates to the host galaxy. In particular, we note that the derived masses are typically one order of magnitude smaller than the inferred O masses in normal core-collapse SNe \citep{Elmhamdi2011, Jerkstrand2015, Dessart2020}.

We caution that elemental abundance estimates at late epochs is challenging with faint emission features. In particular, as the \citet{Uomoto1986} estimate does not capture time evolution, we compare this estimate to detailed models from \citet{Jerkstrand2015} in Figure \ref{fig:oi_model}. As shown, the analytical estimate for the assumed temperature range well constrains the \oi\, luminosity evolution between $\approx 150$ and $\approx 350$\,days (for the nucleosynthetic yields of the \citealt{Jerkstrand2015} models), suggesting that the approximated mass range is a conservative estimate for the total O mass.

The mass estimates derived here are inconsistent with that reported in \citet{Jacobson-Galan2020}, who derive a much higher O mass of $\gtrsim 0.15$\,\Msun. This is likely because i) they derived these estimates using a spectrum at an earlier phase ($\approx 60$\,days from peak) where the source was not completely nebular and ii) they assume that the Ca and O emitting regions are co-located in the ejecta so that the observed \caii/\oi\ ratio directly constraints the Ca/O mass fraction and O mass via the \caii\ luminosity. However, we find this interpretation to be unlikely as detailed modeling of core-collapse SNe has shown that the \caii\ line serves as the primary coolant of the energy deposited in the Si-rich layers, while the \oi\ emission arises from the outer O-rich layers produced largely in the hydrostatic burning phase \citep{Jerkstrand2015, Dessart2020}. Similar arguments for ejecta stratification also have been demonstrated with detailed modeling of thermonuclear shell detonations \citep{Dessart2015a}.

\subsection{Constraints on the progenitor mass}

\begin{figure}
    \centering
    \includegraphics[width=\columnwidth]{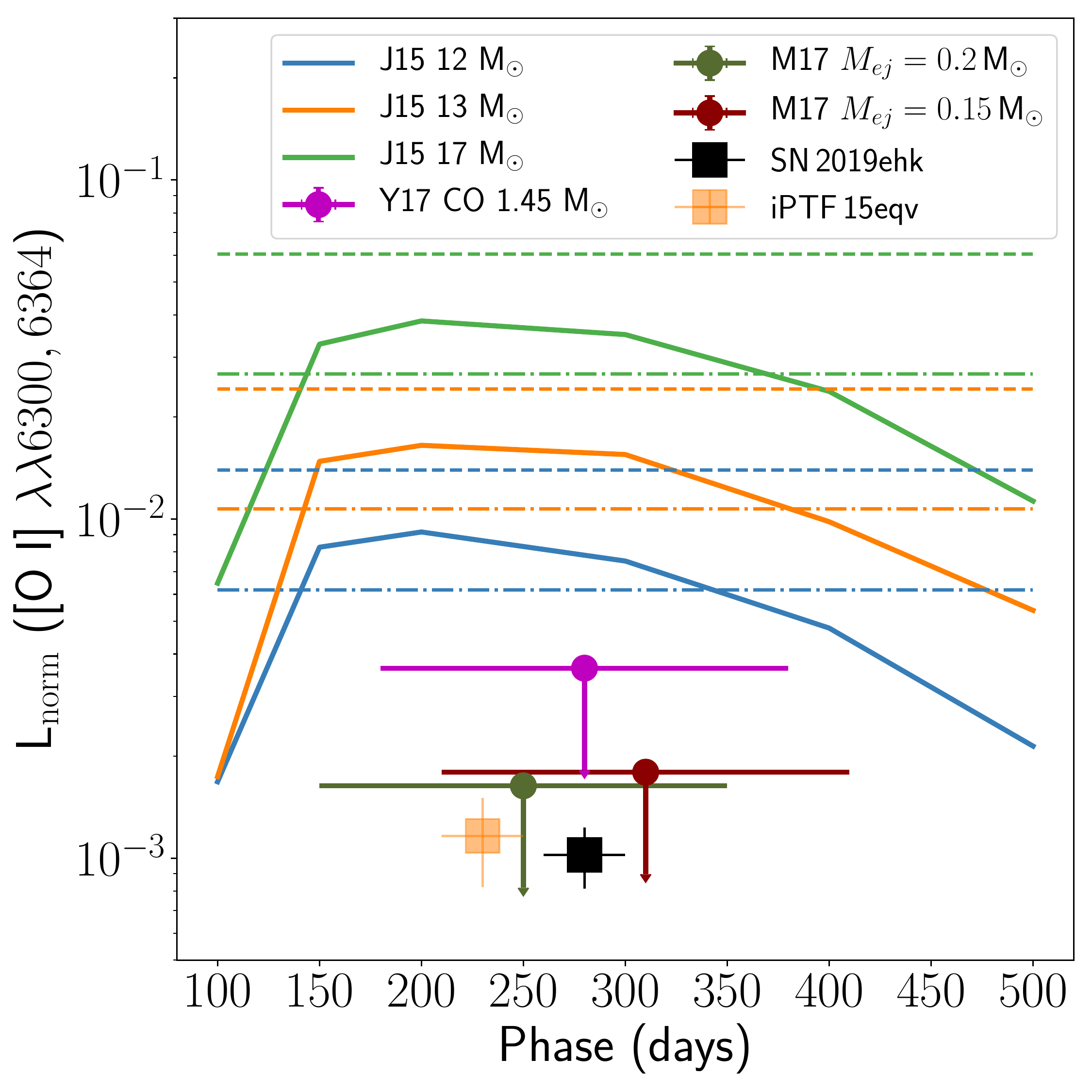}
    \caption{Comparison of the \oi\ luminosity of SN\,2019ehk to models of stripped envelope core-collapse supernovae from \citet[J15]{Jerkstrand2015}. The \oi\ luminosity on the y-axis (denoted as L$_{\rm norm}$) is normalized to the radioactive energy deposition rate from $^{56}$Co decay. We show estimated \oi\ luminosities from the nucleosynthesis calculations of \citet[M17 for different ejecta masses from a 1.5\,\Msun\ CO core]{Moriya2017} and \citet[Y17]{Yoshida2017}, where we use the approximate relationship between \oi\ luminosity and oxygen mass in \citet{Uomoto1986}, assuming a temperature of 3500\,K. The ultra-stripped model luminosity estimates have been arbitrarily shifted in phase for better visualization since the \citet{Uomoto1986} estimate does not capture time evolution. For comparison, we also show the \citet{Uomoto1986} estimate of the \oi\ luminosity for the nucleosynthetic yields of \citet{Jerkstrand2015} and the range of assumed temperatures (between the dot-dashed and dashed horizontal lines of the same color), showing that the time-independent estimates well constrain the \oi\ luminosity evolution between $\approx 150$ and $\approx 350$\,days}. For comparison, we also show the measured normalized \oi\ luminosity of another Ca-rich SN\,IIb iPTF\,15eqv.
    \label{fig:oi_model}
\end{figure}

First, in order to directly compare the observed \oi\ luminosity with detailed nebular phase models of stripped envelope SNe and constrain the progenitor mass, we show in Figure \ref{fig:oi_model}, tracks of the \oi\ luminosity evolution for models of different initial ZAMS masses from \citet{Jerkstrand2015}. As shown, the nebular models of relatively higher mass progenitors ($\approx 12 - 15$\,\Msun) from \citet{Jerkstrand2015} significantly overestimate the \oi\ luminosity, suggesting a much lower progenitor core mass for SN\,2019ehk. Note that this conclusion is independent of the assumed extinction, since the \oi\ luminosity and the $^{56}$Ni luminosity scale similarly with varying extinction. 

Estimates of the O yields for such low progenitor (and CO core) masses are sparse in the literature, and have thus far been calculated for the case of the highly stripped He cores of ultra-stripped SNe \citep{Tauris2013}. In these scenarios, relatively low mass He stars ($\lesssim 3.5$\,\Msun) are stripped down to the CO core by a close binary companion, leaving behind low mass CO cores of $\approx 1.45 - 1.6$\,\Msun\ at the time of explosion \citep{Tauris2015}. Although the presence of strong He lines in the spectra of SN\,2019ehk suggests that the stripping did not extend down to the CO core, the nucleosynthetic O yields in these models are applicable to constrain the CO core mass at the time of explosion. Specifically, we note that the CO core mass is relatively insensitive to the mass loss processes via binary interactions that occur in the very late stages of stellar evolution \citep{Jerkstrand2015, Podsiadlowski1992, Woosley2015, Laplace2020}, and hence a good tracer of the progenitor ZAMS mass \citep{Fransson1989, Jerkstrand2014, Jerkstrand2015}. 

We use the nucleosynthetic yields from \citet{Moriya2017} and \citet{Yoshida2017} to estimate the \oi\ \citep{Uomoto1986} and $^{56}$Ni luminosity in the nebular phase for low mass CO cores of $1.45 - 1.5$\,\Msun, under different assumptions of the explosion energy and ejecta mass. The \oi\, luminosity estimate assumes that all the synthesized O emits in \oi\, and hence serve as upper limits to the observed luminosity.  Figure \ref{fig:oi_model} shows that the upper limits on the \oi\ luminosity for the low mass CO core models are very similar to the low \oi\ luminosity measured for SN\,2019ehk. 

\begin{figure}[!htp]
    \centering
    \includegraphics[width=0.49\textwidth]{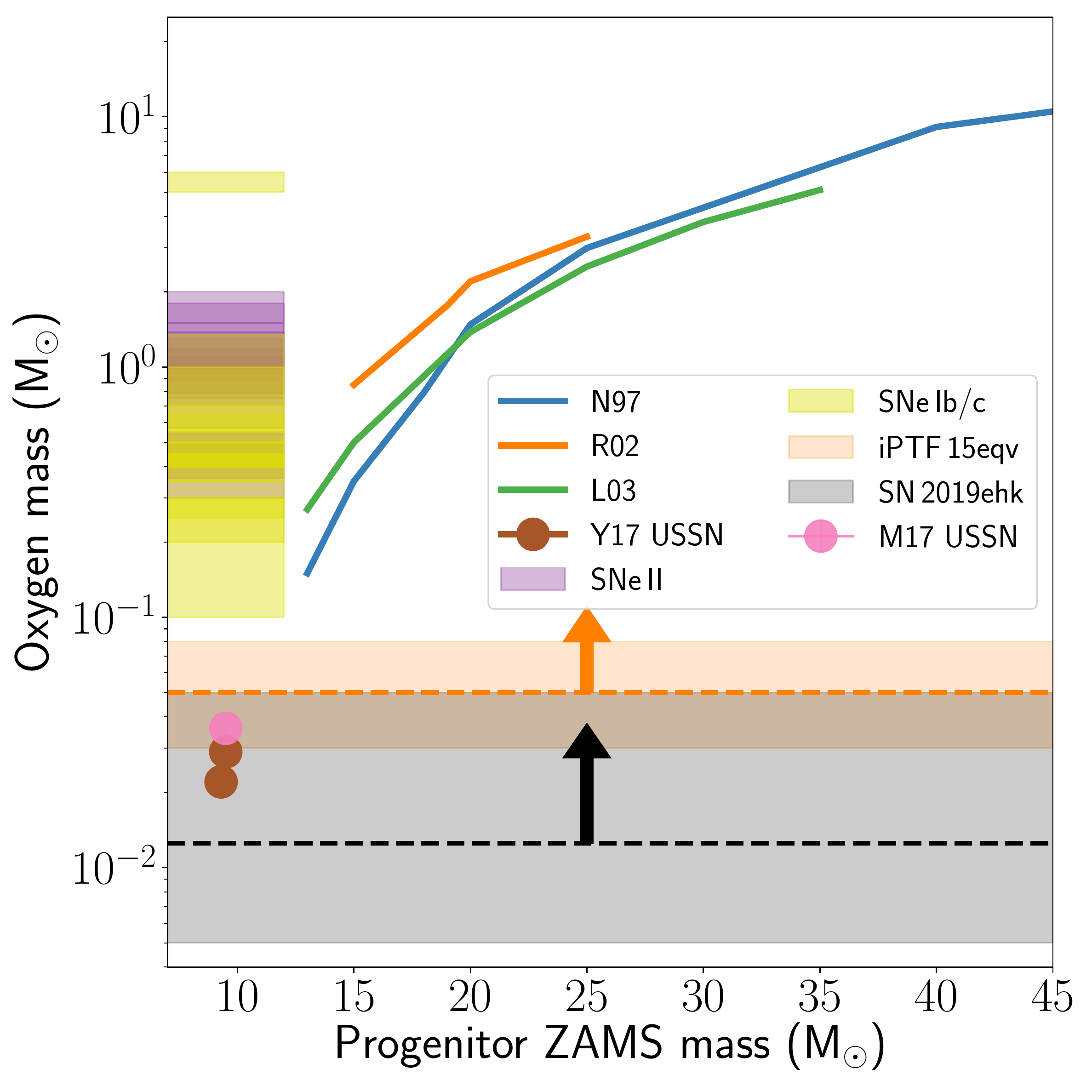}
    \caption{Comparison of the O mass lower limit estimate for SN\,2019ehk (in shaded grey region) to models of synthesized O mass in core-collapse SNe as a function of the progenitor ZAMS mass. The blue, green and orange lines refer to nucleosynthesis models of \citet[N97]{Nomoto1997}, \citet[R02]{Rauscher2002} and \citet[L03]{Limongi2003}. We also plot O nucleosynthetic yields for models of low mass CO cores of ultra-stripped SNe (USSNe) from \citet[M17]{Moriya2017} and \citet[Y17]{Yoshida2017} scaled to the corresponding ZAMS mass expected from stellar evolution \citep{Woosley2015}. For comparison, we plot estimated O masses for normal core-collapse SNe II and Ib/c on the left y-axis. We also show our estimated O mass for a late-time spectrum of another Ca-rich SN\,IIb iPTF\,15eqv.} 
    \label{fig:omass_model}
\end{figure}

Next, we also use the derived O mass limits to constrain the progenitor ZAMS mass for SN\,2019ehk. In Figure \ref{fig:omass_model}, we plot model tracks showing the steep dependence of synthesized O mass on the progenitor ZAMS mass from \citet{Nomoto1997}, \citet{Rauscher2002} and \citet{Limongi2003}.  For comparison we show estimated O masses from a sample of core-collapse SNe of Type II and Type Ib/c from the compilation of \citet{Elmhamdi2011}, demonstrating that the O yields in most normal core-collapse SNe are consistent with $\approx 12 - 20$\,\Msun\ progenitor ZAMS masses. 

Specifically, Figure \ref{fig:omass_model} demonstrates that the small O mass estimated for SN\,2019ehk requires a much smaller progenitor ZAMS mass (and CO core mass) than the canonical models of core-collapse SNe that have been published for ZAMS masses of $\gtrsim 12$\,\Msun\ (corresponding to CO core mass $\gtrsim 2.0$\,\Msun; \citealt{Woosley2015}). We thus compare the O mass estimate to smaller CO core masses that have been simulated in the context of ultra-stripped SNe \citep{Moriya2017, Yoshida2017}. As shown in Figure \ref{fig:omass_model}, the synthesized O mass estimates for these low mass CO cores are consistent with the range estimated for SN\,2019ehk.

\subsection{On the presence of hydrogen in the ejecta}

\begin{figure*}[!htp]
    \centering
    \includegraphics[width=0.8\textwidth]{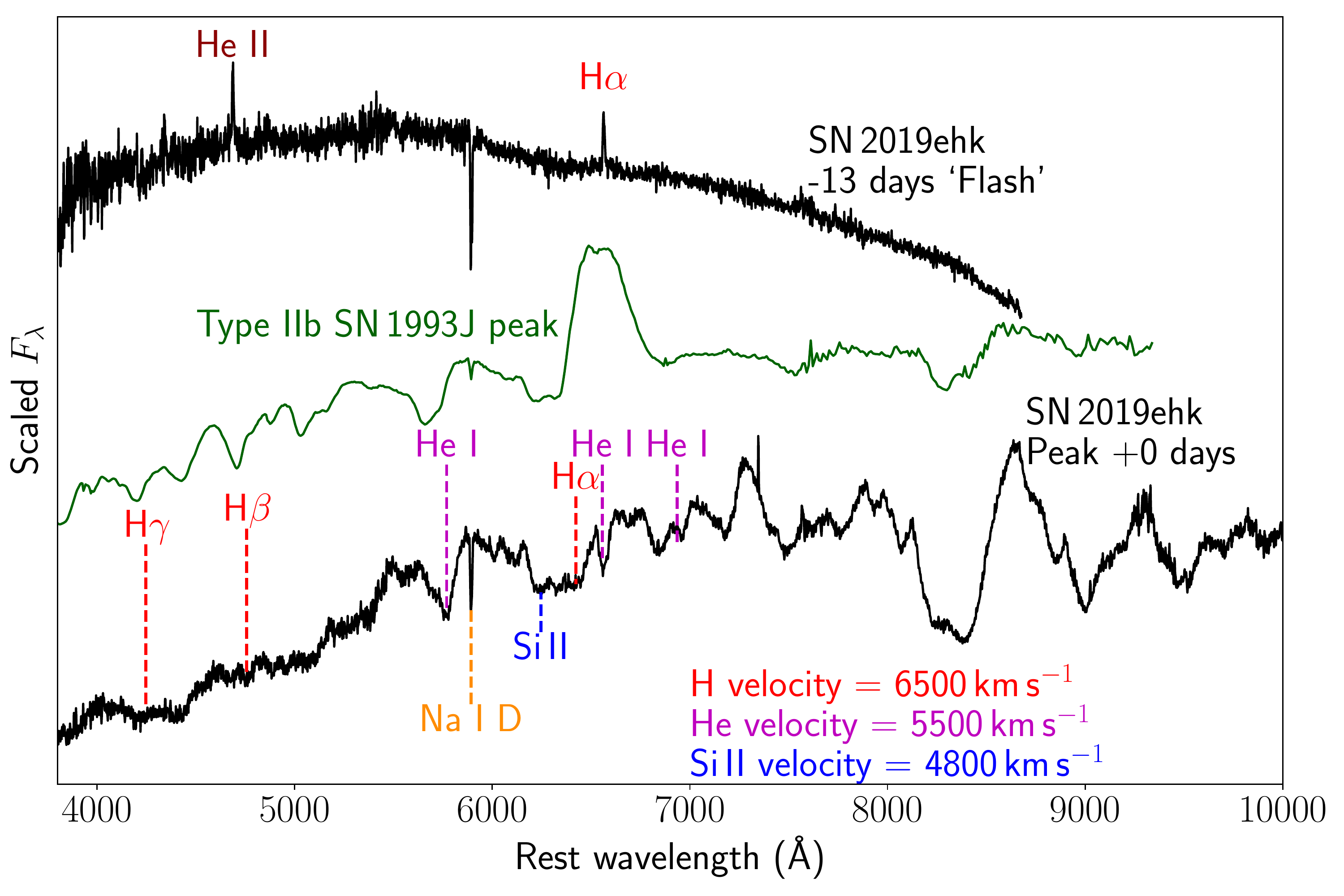}
    \includegraphics[width=0.8\textwidth]{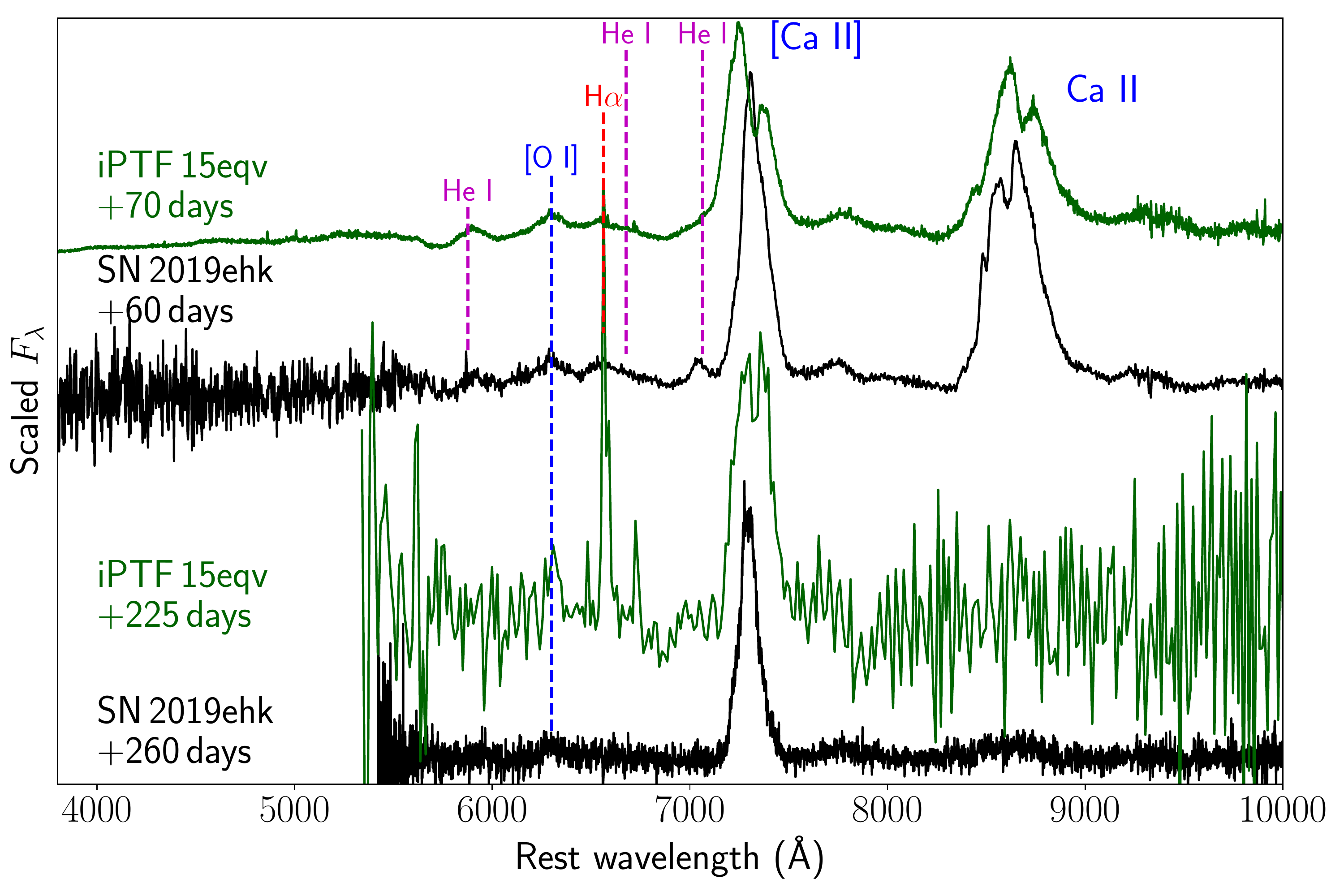}
    \caption{(Top panel) Spectra of SN\,2019ehk around peak light. The very early spectrum at $-13$\,days shows clear narrow emission lines of H$\alpha$ and He II, suggesting a `flash-ionized' hydrogen-rich CSM. We show a comparison of our peak light spectrum of SN\,2019ehk to that of the Type IIb SN\,1993J, highlighting apparent absorption features of H at a velocity of 7500\,km\,s$^{-1}$ and He I at a velocity of 5500\,km\,s$^{-1}$. The spectrum of SN\,1993J has been artificially reddened with $E(B-V) = 0.75$ to match the continuum shape of SN\,2019ehk for better visualization. Note the striking similarities between the two objects in the apparent Balmer and He I absorption features. (Bottom panel) Comparison of the early and late nebular phase spectra of SN\,2019ehk and iPTF\,15eqv (from \citealt{Jacobson-Galan2020} and \citealt{Milisavljevic2017}), highlighting features of H, He I, \oi and \caii.}
    \label{fig:2019ehk_spec}
\end{figure*}

SN\,2019ehk was classified as a hydrogen-poor SN\,Ib in \citet{Jacobson-Galan2020} and \citet{Nakaoka2020}, while \citet{De2020} reported the classification of this object as a Type IIb SN. In Figure \ref{fig:2019ehk_spec}, we plot peak-light optical spectra of SN\,2019ehk together with a spectrum of the Type IIb SN\,1993J \citep{Matheson2000}. We highlight the presence of absorption features in all the Balmer series transitions at velocities of $7500$\,km\,s$^{-1}$, and distinct He I transitions at $5000$\,km\,s$^{-1}$, consistent with compositionally stratified and homologous expanding ejecta for Type IIb SNe \citep{Dessart2011}. We emphasize the similarities between SN\,1993J and SN\,2019ehk in the presence of all the Balmer absorption features as well as the flat-bottomed H$\alpha$ structure seen in other Type IIb SNe \citep{Silverman2009, Marion2014}.

\begin{figure}
    \centering
    \includegraphics[width=\columnwidth]{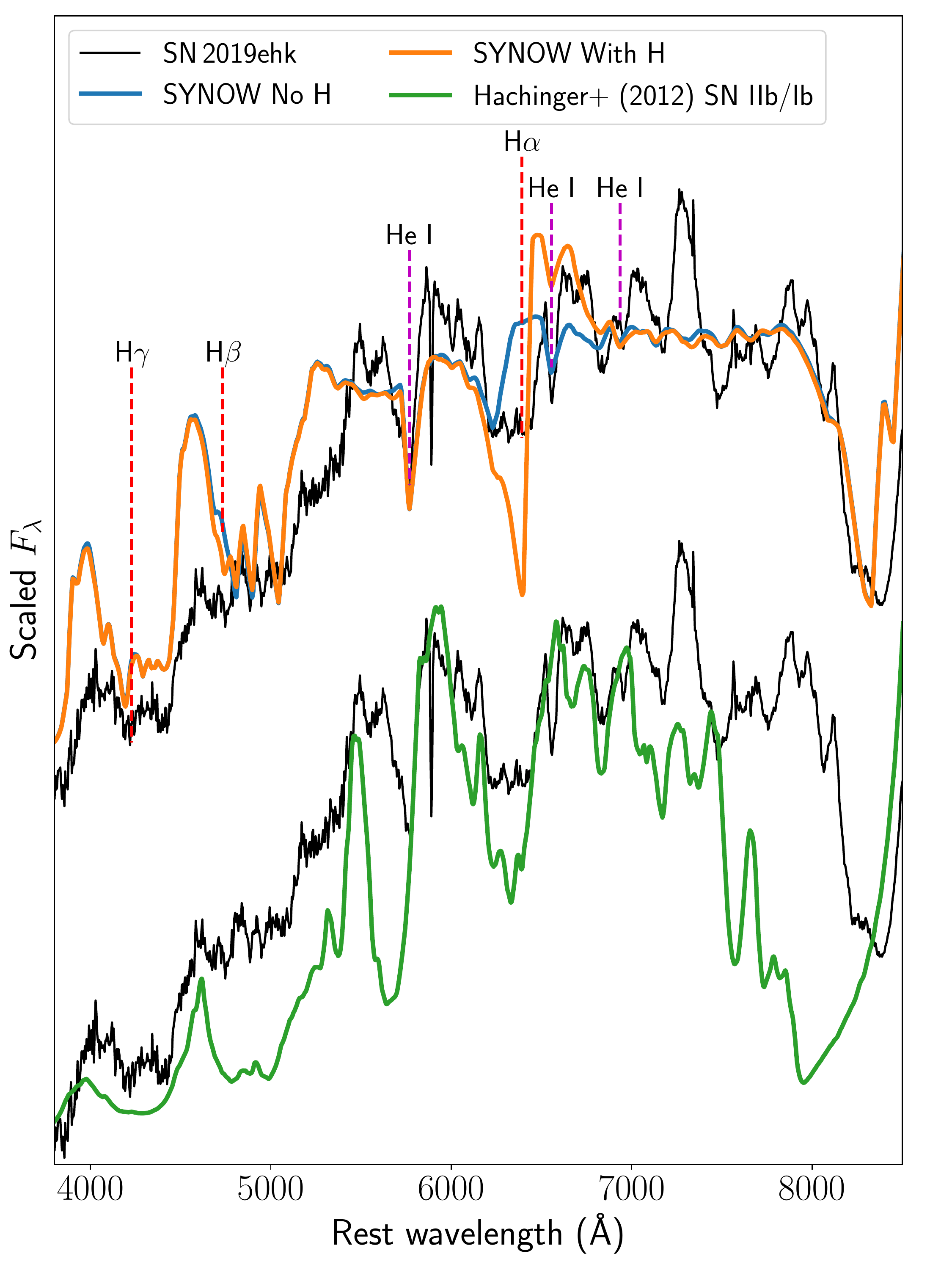}
    \caption{Comparison of the peak spectrum of SN\,2019ehk with a synthetic model created using SYNOW as well as more realistic models for transitional Type Ib/IIb SNe. In the top set of spectra, we overplot the observed spectrum with two SYNOW fits -- one containing hydrogen (in orange) and one without hydrogen (in blue). As shown, the addition of the hydrogen explains the broad trough near $\approx 6500$\,\AA\ as well as the distinctive H$\beta$ feature near $4750$\,\AA. In the lower set of spectra, we overplot SN\,2019ehk (in black) with a transitional SN\,IIb / SN\,Ib model from \citet{Hachinger2012} in green. The model spectra are strikingly consistent with SN\,2019ehk, in particular, explaining the flat-bottomed feature near $\approx 6500$\,\AA\ as well as other Balmer features.}
    \label{fig:modelcompare}
\end{figure}

To demonstrate the presence of hydrogen, we created a synthetic spectrum model for the source using \texttt{SYNOW} \citep{Thomas2011}. We use a combination of the most prominent ions in the observed spectrum at their respective velocities -- He\,I, Si\,II, Ca\,II, Ti\,II and Fe\,II. Using a photospheric temperature of 5000\,K and reddening of $E(B-V) = 0.75$\,mag, we create two spectral models -- one containing H and without H. As shown in Figure \ref{fig:modelcompare}, the combination without H shows a single P-Cygni absorption near $\approx 6100$\,\AA\ from Si\,II but is unable to produce the flat-bottomed feature near $6400$\,\AA. On the other hand, the addition of H explains the shape of that feature as well as the weaker H$\beta$ transition seen around $4750$\,\AA. We note that the \texttt{SYNAPPS} spectroscopic fit (without H) for SN\,2019ehk in \citet{Jacobson-Galan2020} does not reproduce the striking combination of H$\alpha$ and He\,I\,$\lambda6678$ emission/absorption features at $\approx 6400 - 6700$\,\AA, while the H$\gamma$ absorption is not reproduced by their ions. Furthermore, the canonical population of Ca-rich events are not known to exhibit such prominent H$\alpha$ P-Cygni peak and flat bottomed absorption (e.g., Figure 6 in \citealt{De2020}), suggesting SN\,2019ehk is distinctive.

The early time `flash' spectra presented in \citet{Jacobson-Galan2020} also exhibit narrow but resolved emission lines of H$\alpha$ and He\,II (Figure \ref{fig:2019ehk_spec}). Such features are commonly seen in early time spectra of hydrogen-rich core-collapse SNe \citep{GalYam2014, Yaron2017}. In the early nebular phase, SN\,2019ehk exhibits nearly identical spectroscopic features as that of iPTF\,15eqv (Figure \ref{fig:2019ehk_spec}), which was reported as a peculiar hydrogen-rich SN\,IIb which exhibited a nebular phase spectrum dominated by \caii\ emission \citep{Milisavljevic2017}. We specifically note the presence of a broad emission feature near the H$\alpha$ transition, suggesting the presence of H in iPTF\,15eqv. Since H features in Type IIb SNe become weaker with time (as He features get stronger; \citealt{Gal-Yam2017}), the detection of H several weeks after peak light led to the classification of this object as a hydrogen-rich SN\,IIb \citep{Cao2015b, Milisavljevic2017}. At very late phases ($> 200$ days after peak), the spectra both SN\,2019ehk and iPTF\,15eqv are dominated by only \caii\ emission, leading to their classification as `Ca-rich' supernovae.

\citet{Hachinger2012} performed radiative transfer simulations for a range of stripped envelope SN progenitors with varying amounts of H and He left at the time of explosion (see their Figure 10). Although they discuss detailed modeling of SN\,1993J and SN\,1994I, they use their model grid to provide estimates of the amount of H and He required in the ejecta to detect the respective spectral features. Specifically, they show that the flat-bottomed feature near $\approx 6400$\,\AA\ as well as the weaker higher order Balmer series absorption features are commonly seen in their transitional Type IIb/Ib models, formed by absorption from the nearby H$\alpha$ and Si\,II $\lambda6355$ transition. In Figure \ref{fig:modelcompare}, we also compare their transitional SN\,IIb/SN\,Ib model (after reddening) at a phase of $30$ days after explosion\footnote{Given the low ejecta mass (by a factor of $\approx 10$) of SN\,2019ehk compared to normal SNe IIb/Ib/Ic, we expect the optical depth of the ejecta at peak ($\approx 15$\,days after explosion) to be comparable with the models at about $\approx 30$\,days after explosion.} with the observed spectra. The model spectra are strikingly similar in terms of the observed features and explain the clear flat-bottomed feature at $\approx 6500$\,\AA. In particular, although the line ratios are not perfectly reproduced in our \texttt{SYNOW} model, they are consistent with the realistic models presented in this work. These transitional Type IIb/Ib models are achieved with small amounts of residual hydrogen in the progenitor, and suggest a remaining H mass of at least $M_H \approx 0.02 - 0.03$\,\Msun\ in SN\,2019ehk. These estimates are similar to that suggested for other SNe\,Ib possibly showing trace amounts of high velocity hydrogen (e.g. \citealt{Elmhamdi2006}).

\section{Discussion}
\label{sec:discussion}
In this work, we have demonstrated that i) the late-time \oi\ luminosity in SN\,2019ehk is consistent with very low O mass expected for low mass ($\approx 1.45 - 1.5$\,\Msun) CO cores of core-collapse SNe and ii) there is evidence for hydrogen in the early flash-ionized phase, photospheric phase and nebular phase spectra of SN\,2019ehk, suggesting the presence of at least $M_H \approx 0.03$\,\Msun\ in and around the progenitor at the time of explosion. In particular, the presence of photospheric H with multiple transitions at consistent velocities argues for the classification of SN\,2019ehk a SN\,IIb since the SN\,Ib classification has been suggested to be applicable for events that show no H at all \citep{Gal-Yam2017}.

In the case of the interpretation as a thermonuclear transient initiated by a He detonation during a white dwarf merger \citep{Jacobson-Galan2020}, it was suggested that the early time narrow H features were consistent with H-rich CSM (with hydrogen mass of $M_H\sim 10^{-4}$\,\Msun) ejected at the time of merger. However, the presence of photospheric hydrogen suggests $M_H \gtrsim 0.02 - 0.03$\,\Msun, which is difficult to reconcile with this scenario since the progenitor CO + He binary white dwarfs are expected to be very deficient in hydrogen ($M_H \lesssim 10^{-4}$\,\Msun; \citealt{Podsiadlowski2003, Lawlor2006, Zenati2019b}).

While detailed nucleosynthetic yields of the hybrid CO WD merger scenario proposed in \citet{Jacobson-Galan2020} have not been published, we note that the requirement of having $M_H \gtrsim 10^{-3}$\,\Msun based on the early flash spectra was suggested to favor a low mass secondary CO WD of $\approx 0.5$\,\Msun, as more massive WDs have much smaller H layers. On the other hand, it has been shown in previous works that the O yield in sub-Chandrasekhar mass CO core detonations increases rapidly for smaller core masses \citep{Sim2010, Townsley2016, Shen2018, Polin2019a}. For instance, detonations of $0.8$\,\Msun cores in \citet{Shen2018} produce $\approx 0.2 - 0.3$\,\Msun of O while $1.0$\,\Msun cores produce $\approx 0.05 - 0.10$\,\Msun of O. Similarly, detonation of the lowest mass $0.6$\,\Msun cores in \citet{Polin2019a} produce $\approx 0.48$\,\Msun of mostly O. We thus find the requirement of substantial amounts of residual H together with the small O yield inferred from the data are inconsistent with the WD scenario.

While \citet{Nakaoka2020} suggested that SN\,2019ehk originated in an ultra-stripped core-collapse SN, hydrogen in not expected in the ejecta of ultra-stripped SNe with compact objects as close binary companions \citep{Tauris2015}. However, low mass progenitors of stripped core-collapse SNe can retain a large range of H and He masses depending on the nature of the companion and the initial binary period \citep{Yoon2010, Zapartas2017, Laplace2020}. The O mass estimate for SN\,2019ehk suggests a ZAMS $\approx 9 - 9.5$\,\Msun\ progenitor that forms a He core mass of $\approx 1.8 - 2.0$\,\Msun. Assuming a residual H mass of $\lesssim 0.1$\,\Msun, the inferred ejecta mass of SN\,2019ehk of $\approx 0.5 - 0.6$\,\Msun\ is consistent with a final CO core mass of $\approx 1.45 - 1.5$\,\Msun\, that collapses to form a $\approx 1.3$\,\Msun\ neutron star and ejects $\approx 0.5$\,\Msun\ of material.

The evidence for dense nearby CSM as seen in the early time light curve and spectra \citep{Jacobson-Galan2020} would then be explained by elevated mass loss prior to explosion as expected for low mass He cores \citep{Woosley2019, Laplace2020}. Comparing our inferred parameters of SN\,2019ehk to the single star models of \citet{Woosley2015} and binary models of \citet{Laplace2020}, who present detailed calculations of the late phase evolution of low mass He cores, we find that their solar metallicity models of progenitors between $9.0$ and $9.5$\,\Msun\ are strikingly similar to the estimated CO core mass and large pre-explosion radius (see Table A.2 in \citet{Laplace2020}). Finally, the low mass stripped core-collapse progenitor scenario is consistent with the $< 10$\,\Msun\ star pre-explosion imaging constraints discussed in \citet{Jacobson-Galan2020}.

Recently, \citet{Jacobson-Galan2020b} presented additional late-time photometry of SN\,2019ehk out to $\approx 390$\,days from peak, and presented two primary arguments against the massive star scenario. First, they estimate the O mass in the ejecta to be $\approx 0.2 - 0.35$\,\Msun, much larger than our estimates. However, their estimate is derived assuming that the Ca abundance in the ejecta is very small based on the arguments in \citet{Jacobson-Galan2020}. As we argue in Section \ref{sec:results}, their Ca mass ratio limits are derived assuming nearly complete mixing of the Ca and O regions, which has been demonstrated to be unlikely in both core-collapse and thermonuclear SNe. Thus, if the Ca contributes substantially more opacity in the ejecta, then the required O mass would be much smaller. Next, they compare their estimate of the $^{57}$Co/$^{56}$Co mass ratio in the ejecta to different progenitor models. In particular, we note that their $^{57}$Co/$^{56}$Co estimate is similar to that expected for the low mass CCSN models (green hexagons in their Figure 4) in \citet{Wanajo2018}, except with a larger ejecta mass. Since the ejecta mass depends sensitively on the nature and separation of the companion in the final stages prior to core-collapse, while the nucleosynthetic yields are largely unaffected, we find that our favored model of a low mass CCSN remains consistent with their measurements.

The interpretation of the Ca-rich SN\,2019ehk as a core-collapse SN adds another member to a growing class of core-collapse SNe that exhibit strong \caii\ lines\footnote{In the case of the fast evolving ultra-stripped SN\,2019dge \citep{Yao2020}, the late-time spectrum was dominated by CSM interaction with He-rich material, likely hiding the underlying nebular emission features from the ejecta} -- the others being the SN\,IIb iPTF\,15eqv \citep{Milisavljevic2017}, SN\,Ic iPTF\,14gqr \citep{De2018a}, and possibly the SN\,Ib iPTF\,16hgs \citep{De2018b}, although iPTF\,16hgs may also be consistent with a thermonuclear detonation. \citet{Kawabata2010} suggested that the SN\,Ib 2005cz could also have originated via this scenario, although its old environment argues against this interpretation \citep{Perets2011}. Some of the Ca-rich SNe reported in \citet{Fillipenko2003} that were found in star forming galaxies may also be members of this class, although their poor photometric and spectroscopic coverage precludes a secure identification \citep{Kasliwal2012a}.

iPTF\,15eqv is perhaps the closest analog of SN\,2019ehk, and was also shown to be a Type IIb core-collapse SN in a star forming environment \citep{Milisavljevic2017} with a large \caii/\oi\ $\gtrsim 10$. For comparison, as in SN\,2019ehk, we calibrate the latest nebular spectrum of iPTF\,15eqv at $\approx 225$\,days with reported late-time photometry to derive the \oi\ luminosity with \caii/\oi\ = 10 \citep{Milisavljevic2017}. We derive a \oi\ luminosity of $\approx 1.2 \times 10^{38}$\,erg\,s$^{-1}$, corresponding to a O mass of $\approx 0.03 - 0.08$\,\Msun\ (shown in Figure \ref{fig:omass_model}). Using the estimated range of $^{56}$Ni masses for this object, we also plot the normalized \oi\ luminosity for this object in Figure \ref{fig:oi_model}. Both the \oi\ luminosity and the O mass estimate for this object is consistent with a very low mass progenitor similar to SN\,2019ehk. 

Taking the large nebular \caii/\oi\ ratio and low \oi\ luminosity as a signature of the low progenitor mass, the primary difference between SN\,2019ehk and iPTF\,15eqv would then be the final mass at the time of explosion. This leads to the different ejecta masses of $\approx 0.5$\,\Msun\ in SN\,2019ehk and $\approx 2 - 4$\,\Msun\ in iPTF\,15eqv \citep{Milisavljevic2017}, where iPTF\,15eqv has a slightly more massive O core (higher \oi\ luminosity) and H envelope (larger ejecta mass). Since stars in this low mass range ($\approx 9 - 9.5$\,\Msun) are still left with massive H envelopes of $\approx 7$\,\Msun\ at the time of the SN \citep{Woosley2015} in single star evolution, the differences between the progenitors can be explained as differences in the binary stripping, which is a function of the nature and orbital period of the companion.

\section{Summary}
\label{sec:summary}
We have presented very late-time imaging of the peculiar Ca-rich SN\,2019ehk with the Keck-I telescope, which we use to perform accurate flux calibration of a contemporaneous late-time spectrum, and derive fluxes for the two most prominent nebular phase lines of \oi\ and \caii. In addition, we presented a high signal-to-noise peak light optical spectrum of the source, which we use to constrain the ejecta composition. To summarize our findings,

\begin{itemize}
    \item The low \oi\ luminosity in the nebular spectrum of SN\,2019ehk suggests a very low O mass of $\approx 0.004 - 0.069$\,\Msun\ (over the range of extinction and temperature assumptions). The inferred value is at least one order of magnitude smaller than that inferred for typical SNe II and SNe Ib/c.
    \item Comparing the inferred O mass to models of core-collapse SNe, we find consistency with the O yields expected from low CO cores of $\approx 1.45 - 1.5$\,\Msun, corresponding to He core masses of $\approx 1.8 - 2.0$\,\Msun\ and ZAMS masses in the range of $\approx 9.0 - 9.5$\,\Msun, as derived from models of massive stars in both single and binary systems. 
    \item We highlight the presence of Balmer series features in the peak light and early nebular phase spectra of SN\,2019ehk, as well as the striking similarity of the H$\alpha$ profile shape to previous observations and radiative transfer models of SNe\,IIb. In addition, the H-rich CSM inferred from very early photometry and spectroscopy is similar to that observed in several young Type II core-collapse SNe. We thus suggest the classification of SN\,2019ehk as a SN\,IIb.
    \item We find that the presence of photospheric hydrogen features (suggesting $M_H \gtrsim 0.02 - 0.03$\,\Msun) is inconsistent with models involving the thermonuclear detonation of a He shell during a white dwarf merger, as they are expected to retain only $M_H \sim 10^{-4}$\,\Msun.
    \item We thus favor the interpretation of SN\,2019ehk as a core-collapse supernova from a low mass $\approx 9.5$\,\Msun\ progenitor, which has been stripped of most of its hydrogen envelope by a binary companion.
\end{itemize}

Our results provide evidence for a class of Ca-rich core-collapse SNe (including SN\,2019ehk and iPTF\,15eqv) from low mass CO cores that form a distinct population from the thermonuclear Ca-rich gap transients found in old environments. While it is currently not obvious what photometric and spectroscopic properties distinguish this class from the old thermonuclear Ca-rich transients (apart from their star forming host environments), the presence of hydrogen in the ejecta of some objects (as demonstrated by peak light spectra of iPTF\,15eqv and SN\,2019ehk) provides strong evidence for the massive star scenario where the progenitors can retain a substantial amount of hydrogen ($M_H \gtrsim 0.01$\,\Msun) at the time of explosion. Detailed nebular phase modeling of the nucleosynthetic products generated from core-collapse explosions of low mass CO cores, which have not been presented in the literature till this date, hold the potential to reveal significant insights into this phenomenon.

\acknowledgements
We thank S. Hachinger for providing the model spectral sequences used in this work. We thank the anonymous referee for a careful reading of the manuscript that significantly improved the quality of the paper. We thank W. Jacobson-Galan for providing the spectral sequence of SN\,2019ehk. We thank J. Sollerman and C. Fransson for constructive comments on this manuscript. We thank L. Bildsten and A. Polin for valuable discussions. We thank D. Perley for assistance with \texttt{lpipe}.

M.~M.~K. acknowledges generous support from the David and Lucille Packard Foundation. This work was supported by the GROWTH (Global Relay of Observatories Watching Transients Happen) project funded by the National Science Foundation under PIRE Grant No 1545949. Some of the data presented herein were obtained at the W.M. Keck Observatory, which is operated as a scientific partnership among the California Institute of Technology, the University of California and the National Aeronautics and Space Administration. The Observatory was made possible by the generous financial support of the W.M. Keck Foundation. The authors wish to recognize and acknowledge the very significant cultural role and reverence that the summit of Mauna Kea has always had within the indigenous Hawaiian community. We are most fortunate to have the opportunity to conduct observations from this mountain. AGY’s research is supported by the EU via ERC grant No. 725161, the ISF GW excellence center, an IMOS space infrastructure grant and BSF/Transformative and GIF grants, as well as The Benoziyo Endowment Fund for the Advancement of Science, the Deloro Institute for Advanced Research in Space and Optics, The Veronika A. Rabl Physics Discretionary Fund, Paul and Tina Gardner, Yeda-Sela and the WIS-CIT joint research grant;  AGY is the recipient of the Helen and Martin Kimmel Award for Innovative Investigation.

\end{document}